%% file: misiriotis.tex
\documentclass{aa}
\usepackage{txfonts}
\usepackage{graphicx}           

\begin{document}

\title{Dust masses and star formation in bright IRAS galaxies}
\subtitle{Application of a physical model for the interpretation of FIR observations}
\author{A. Misiriotis \inst{1}
\and I. E. Papadakis \inst{1,2}
\and N. D. Kylafis \inst{1,2}
\and J. Papamastorakis \inst{1,2}
}

\offprints{Angelos Misiriotis, (\email{angmis@physics.uoc.gr})}

\institute{University of Crete, Physics Department, P.O. Box 2208, 710 03
Heraklion, Crete, Greece
\and Foundation for Research and Technology-Hellas, P.O. Box 1527, 711 10 Heraklion, Crete, Greece}

\date{Received / Accepted}
\abstract{We address the problem of modeling the far-infrared (FIR)  
spectrum and deriving the star-formation rate (SFR) and the dust mass
of spiral galaxies. We use the realistic physical model of Popescu et al.
(\cite{popescu}) to describe the overall ultra-violet (UV), optical
and FIR spectral energy distribution (SED) of a spiral galaxy.
The model takes into account the 3-dimensional old and
young stellar distributions in the bulge and the disk of a galaxy, together with
the dust geometry. The geometrical characteristics of the galaxy and the intrinsic
optical and near-infrared spectra are determined by the galaxy's observed
K-band photometry. The UV part of the spectrum is assumed to be proportional
to the SFR through the use of population synthesis models. By solving the
radiative transfer equation, we are able to determine the absorbed energy,
the dust temperature and the resulting FIR spectrum. The model has only
three free parameters: SFR, dust mass, and the fraction of the UV
radiation which is absorbed locally by dense dust in the HII regions. 
Using this model, we are able to fit well the FIR spectra of 62
bright, IRAS galaxies from the ``SCUBA Local Universe Galaxy Survey" of
Dunne et al.(\cite{dunne1}). As a result, we are able to determine, among others, their
SFR and dust mass. We find that, on average, the SFR (in absolute units),
the star-formation efficiency, the SFR surface density, and the ratio of FIR
luminosity over the total intrinsic luminosity, are larger than the
respective values of typical spiral galaxies of the same morphological
type. We also find that the mean gas-to-dust mass ratio is close to the
Galactic value, while the average central face-on
optical depth of these galaxies in the V band is 2.3. 
Finally, we find a strong correlation between
SFR or dust mass and observed FIR quantities like total FIR luminosity or FIR
luminosity at 100 and 850 $\mathrm{\mu m}$. These correlations yield well defined
relations, which can be used to determine a spiral galaxy's SFR
and dust-mass content from FIR observations.
}
\maketitle

\section{Introduction}

The galactic environment is a complicated mixture of stars, gas and dust.
Stars originate from gas while the presence of dust catalyzes the process.
In turn the dust absorbs the light of the stars (new-born and old) to emit the
absorbed energy in the far infrared. The relation between star formation,
gas and dust is of indispensable importance in any attempt to understand
the evolution of galaxies (see Calzetti \cite{calzetti} for a review on
the subject). Thus, a reliable method to determine the SFR has been among
the main goals of research the last few years. A plethora of diagnostics
have been introduced (see Kennicutt \cite{kennicutt2} for a review) to
address the problem of deriving an estimate for the SFR of a given galaxy.
One of the main problems of most methods is the attenuation from dust that
obscures the measured quantities in the UV and optical bands.  For example,
Charlot et al. (\cite{charlot}), using a combination of population
synthesis models and a photoionization code, concluded that standard
estimators based on $\mathrm{H\alpha}$ and $\mathrm{[OII]}$ underestimate the SFR by
a factor of ~3. On the other hand, they find that their SFR estimates
agree very well with estimates based on the 60 and 100 $\mathrm{\mu m}$ fluxes,
indicating that a promising way to overcome the problem of attenuation of
traditional SFR diagnostics is the use of dust emission itself as a SFR
diagnostic. This is straightforward in starburst galaxies where the dust
opacity is high and most of the dust heating originates from the young
stellar population. In this case, the FIR luminosity can be used as a
direct measure of the SFR (e.g. Lehnert \& Heckman \cite{lehnert}; Meurer
et al. \cite{meurer1})

However, the situation is more complicated in normal spiral galaxies where
the stellar luminosity is not completely reprocessed by interstellar dust.
For example, Xu \& Buat (\cite{xu}) compared the UV, optical and FIR
emission (estimated from IRAS measurements)
from a sample of 135 spirals and found that only 30\% of the
total bolometric luminosity is emitted at the FIR.
A more recent work, based on a direct measurement of the bulk
of the dust emission
\footnote{using ISO data from Tuffs et al. (\cite{tuffs})}
was obtained by Popescu \& Tuffs and
(\cite{popescu2}) suggests that the percentage of stellar light 
re-radiated by dust lies between 15 to 50\%. 

These studies indicate that the moderate optical
depth of normal galaxies allows a significant fraction of the UV radiation
to escape the galaxy without heating the dust. Furthermore, a percentage
of the dust heating in normal galaxies can be attributed to the old
stellar population as well. These two effects complicate the relation
between SFR and FIR in these galaxies.

To address properly the relation between SFR and FIR in normal galaxies
with moderate optical depth the use of a detailed theoretical model for
the FIR emission is required. The origin of the dust heating, the dust
mass, and the optical depth are questions entangled in the problem of
defining the SFR-FIR relation, and should be addressed simultaneously by
the model. The model should also take into account the geometry of the
stars (both old and young) and the dust. The radiative transfer equation (RTE)
should be solved and the emerging spectrum from the UV to the FIR should
be calculated. Comparison of the theoretical SEDs with observed FIR
spectra will then allow an accurate investigation of whether and how the
various observational properties (i.e. FIR luminosity, flux at certain
bands, FIR colors) correlate with the fundamental physical parameters like
SFR and dust mass.

A few authors have followed this approach in the past. For example, Buat
\& Xu (\cite{buat}) (see also Buat et al. \cite{buat2}) compared the UV
with the FIR emission of 152 disk galaxies.  They used a simplified model
to calculate the extinction at the UV band. Given the extinction they
corrected the UV emission and calibrated the FIR luminosity in terms of
SFR. In the same fashion, Hirashita et al. (\cite{hirashita}) derived
extinction correction for several SFR indicators based on the simple
assumption that all the estimators should return the same SFR. They
applied their method on a sample of 47 irregulars and spirals and 32
starburst galaxies and emphasized the role of dust extinction on several
SFR tracers. Bianchi et al. (\cite{bianchi1}) used a detailed Monte Carlo
radiative transfer code and applied it to one object, namely
NGC~6946.  They used this code to model its overall SED, and concluded
that it's central face-on optical depth is of the order of 5.  According
to the predictions of their model, the old stellar population is the main
source of dust heating and therefore they draw no conclusions on the SFR of
this galaxy. The intrinsic colors of a stellar population depend on the
star-formation history of this population. Therefore several authors (e.g.
Silva et al. \cite{silva}; Devriendt et al. \cite{devriendt}; Efstathiou
\& Rowan-Robinson \cite{efstathiou}) have introduced population synthesis
models to describe the intrinsic SEDs of their models and subsequently
applied extinction and emission from dust to derive the emerging  SED. 
Yet, these models tend to involve too many parameters that can not 
be strictly constrained.

Finally, Popescu et. al (\cite{popescu}; P00 hereafter) developed the detailed theoretical
model which we use in this paper and which was first tested and applied
to the edge-on spiral NGC~891. In their work, they used a detailed
geometric description of the galaxy based on optical observations from
Xilouris et al. (\cite{xilouris2}). They also included the effect of
localized absorption of the UV in star forming regions. They found that
30\% of the dust heating is attributed to the old stellar
population. The model of P00 was successful not
only in fitting the SED of NGC~891 but also in predicting the
FIR morphology of this galaxy (Popescu et al. {\cite{popescu3}).

Furthermore, the same model was applied on 4 more edge-on galaxies by
Misiriotis et al. (\cite{misiriotis}) to derive their SFR and dust mass.
Their results are in excellent agreement with the SFRs derived by
Kennicutt (\cite{kennicutt1}) from a sample of 61 normal spiral galaxies.

Since FIR observations that cover the full range from 10 to 1000 $\mathrm{\mu m}$
are becoming rapidly available, it is desirable to develop a method which
will allow us to use these observations to derive important physical
parameters of spiral galaxies, such as the SFR and the dust mass. In this
paper we present the first application of the P00 model
to a large number of spiral galaxies with IRAS data and recent
high quality FIR measurements (Dunne et al. \cite{dunne1}; DU00
hereafter). The use of a realistic physical model and the excellent
agreement between the model SEDs and the FIR measurements, allow us to
derive reliable SFR and dust-mass values. Furthermore, the large number
of galaxies considered in this work allows us to detect significant
correlations and derive important diagnostic relations between various
observational FIR emission properties and these physical parameters.

In Sect.~\ref{sec:model} we discuss the physical model for the FIR
emission that we are using. In Sect.~\ref{sec:sample} we present the data
for the sample of galaxies we consider in this study, and in
Sect.~\ref{sec:fiting} we describe the fitting procedure of our model SEDs
to the observed FIR spectra. In Sect.~\ref{sec:results} we present and
discuss the results from the model fitting procedure, namely the SFRs and
the dust masses for the galaxies in our sample. In
Sect.~\ref{sec:diagnostics} we show diagnostic relations that we derived
between measured FIR and physical quantities, and we summarize in
Sect.~\ref{sec:summary}.

\section{Description of the FIR emission model}
\label{sec:model}

The model used in this work is based on the model of P00, where the reader
should refer for more details. Two simplifications
were made due to the statistical character of this study.
\begin{enumerate}
\item
In P00 the dust is distributed in two exponential disks with different
scalelengths and scaleheights. This detailed approach was directed by the
large amount of data available for NGC~891. In this study such data are
not available and therefore only one exponential disk is used to describe
the dust distribution.
\item
In P00 the stochastic heating of the diffuse dust was taken into account.
In the present work, the stochastic heating was neglected due to the
computational load it would require.
\end{enumerate}

In the rest of this section we will present an outline of the model and 
discuss the assumptions we made due to the lack of detailed data for our sample.
Our primary aim is to construct a theoretical model which will allow us to
reproduce reliably the overall UV, optical and FIR SED emerging from a spiral
galaxy. To this end, one has to determine the intrinsic luminosity
of a galaxy in several wavelengths as well as the parameters describing
the spatial distribution of the stars (e.g. the disk and bulge
scalelengths, scaleheights etc.) and dust. Then, given the stellar and dust
spatial distributions, one has to solve the RTE. The solution of the RTE
provides among others the absorbed energy that heats the dust. At the same
time, the extinction effects in the UV and the optical part of the SED are
calculated. According to the dust's temperature and emission properties,
the absorbed energy is re-emitted in the FIR. Since the absorbed energy
varies from point to point, the RTE must be
solved for a grid within the model galaxy. Then, the dust emission from
all the grid points is integrated to get the total emission spectrum of
the dust.

For edge-on spiral galaxies, the parameters which determine the geometry
of stars and dust can be found using the method introduced by Kylafis \&
Bahcall (\cite{kylafis}) and subsequently used by Ohta \& Kodaira (\cite{ohta}),
 Xilouris et al. (\cite{xilouris1}; \cite{xilouris2}; \cite{xilouris3}),
 Kuchinski et al. (\cite{kuchinski}). This method consists of
finding a 3D distribution for the stars and the dust that replicates the
actual observed image of the galaxy. The prominent dust lane in edge-on
galaxies constrains the dust mass as well as it's scalelength and scaleheight.

However, the observational determination of all the photometric and
geometric parameters is not possible for the majority of the spiral
galaxies, which are not seen edge-on. In this case, one has to assume the
3D stellar and dust distributions, based on the results from 2D bulge-disk
decomposition studies. In our case we choose to represent a spiral galaxy
using an exponential stellar disk, an oblate exponential bulge with axial ratio 0.5
and an exponential dust disk.

\subsection{Spatial distribution of the old stellar population}
In order to determine the geometric parameters of the stellar distribution in
a galaxy we use mainly the results of de Jong (1996; DJ96 hereafter),
which are based on optical and near-infrared observations of 86 face-on
spiral galaxies. 
We start by determining the scalelength, $h_s$ (in pc), of the disk (which
represents the old stellar population) with the use of the relation:
$M_{K,d}=-6-5\log(h_s)$, where $M_{K,d}$ is the K-band absolute magnitude
of the disk. This relation was determined from the ``K-band disk central
surface brightness" versus ``disk scalelength" relation of DJ96 using only
those galaxies with morphological type T lying within $0\le T \le 6$.
Similar results have been reported by M\"{o}llenhoff \& Heidt
(\cite{mollenhoff}) and Graham (\cite{graham}). Given the radial
scalelength of the disk, it's scale-height $z_s$ is calculated by
$z_s=0.25 h_s$ (Pohlen et al. \cite{pohlen}). As for the geometry of
the stellar population in the bulge, we assume that its scalelength,
$h_b$, is equal to $0.14 h_s$ (DJ96). The K-band absolute magnitude
of the bulge, $M_{K,b}$, is determined by $M_{K,d}$ through the relation:
$M_{K,b}=16.5+1.6 M_{K,d}$, which is the result of a linear fit
to the DJ96 results (their Fig. 19) considering galaxies with $0\le T
\le 6$ only.

\subsection{Dust spatial distribution and emission properties}
The diffuse dust is distributed in another exponential disk with
scalelength $h_d$ and scaleheight $z_d$ given by
$h_d=1.4 h_s$ and $z_d=z_s/1.8$ (Xilouris et. al \cite{xilouris3}).
For the dust absorption and emission properties we are using the Milky-Way
$R_V=3.1$ model of Weingartner \& Draine (\cite{weingartner}).

Apart from the diffuse dust, a small but warm ammount of dust
is expected to be found in HII regions.
As in P00, a fraction F of the UV originating  from the young stellar
population is absorbed locally by the dust in these regions. 
In this study, we assume that this locally absorbed energy is re-emitted according
to a spectrum template typical for star forming regions. The template is
based on IRAS observations of HII regions in the Galaxy (Chini et
al. \cite{chini}; Chan \& Fich \cite{chan})  and in the Large Magellanic
Cloud (Bell et al. \cite{bell}). In effect, this template is well
approximated by thermal emission from dust with temperature $35\degr\mathrm{K}$.

In order to reduce the computational time of the model we did not take into
account the stochastic heating of the diffuse dust. The dust emission due to
stochastic heating affects the diffuse dust emission at 60 and 100$\mathrm{\mu m}$
but the effect is small compared to the contibution from the localized
dust emission from HII regions in these wavelengths. However, one expects
that the inclusion of the stochastic heating would systematically yield
more radiation from the diffuse dust at 60 and 100$\mathrm{\mu m}$.
As a result less room would be left for dust emisson from the HII regions
and the value of F would be lower.

\subsection{Intrinsic luminosity of the old stellar population}
Under these assumptions, the overall size and the bulge-to-disk ratio of each galaxy
are essentially determined by $M_{K,d}$ and its morphological type. 
Furthermore, the disk and bulge magnitudes in the K band
determine the intrinsic optical and near-infrared luminosity of 
each galaxy in the B, V, R, I, J and H bands through the relations: 
B-V=0.78, B-R=1.23, B-I=1.76, B-H=3.30 and B-K=3.59.  
These are the weighted mean integrated colors
for spiral galaxies with $0\le T\le 6$, which we use for both the disk and
bulge, since they are very similar (DJ96).  Finally, we adopt B-J=2.69
as given by M\"{o}llenhoff \& Heidt (\cite{mollenhoff}), based on their
results from the surface photometry of 40 face-on disk galaxies. Note that
the choice of the K-band luminosity as the basis for the determination of
the intrinsic UV/optical radiation field was motivated by the fact that
the light emitted in this band is the least affected by dust absorption
when compared to the light emitted in other frequently observed bands,
i.e. B through H. As a result, the observed K-band magnitudes can
be considered as an accurate measure of the intrinsic galaxy luminosity, 
in this band.

\subsection{The young stellar population}
Finally, for the young stellar population we assume that it's spatial
distribution can be described by an exponential disk with the same
scalelength and scaleheight as that for the dust.  The young stellar disk's
luminosity is proportional to the SFR. Thus the luminosity of the young
stellar population, $L_{\lambda}$, is expressed in terms of the SFR using
the relations $SFR=8.12\times 10^{-28}L_{\lambda}\mathrm{[erg\, s^{-1}\, Hz^{-1}]}$ at $\lambda=912$\AA\
and $SFR=1.4\times 10^{-28}\,L_{\lambda}\mathrm{[erg\, s^{-1}\, Hz^{-1}]}$ at $\lambda=1200$\AA\ and
$\lambda=2800$\AA.
These values were calculated with the PEGASE
population synthesis model of Fioc \& Rocca-Volmerange (\cite{fioc})
assuming a Salpeter initial mass function with mass limits from 0.1 to
100$M_\odot$, constant SFR and metalicity Z=0.02. Note that different
authors adopt slightly different conversion factors, depending on the
parameters of the adopted population synthesis model (e.g. Bruzual \&
Charlot \cite{bruzual}).
The intrinsic luminosity at the U band is attributed both to the young and the
old stellar populations. Therefore, we have set it to the average of $L_{2800}$ 
(which is assumed to originate only from the young stellar population) and $L_B$ 
(which is assumed to originate only from the old stellar population).
The ionizing UV (shortwards 912\AA) was neglegted since we assumed that it does not
contribute significantly to dust heating

Details on the approximations we follow to solve the RTE can be found in 
Kylafis \& Bahcall (\cite{kylafis}).
Since the geometry of the old and young stellar population and
dust distribution, as well as the intrinsic old population luminosity in
the optical and near-infrared bands are determined by the morphological
type and by the disk and bulge K-band magnitudes, the theoretical
SED is determined by introducing three free parameters
only:

\begin{enumerate}

\item
The central face-on optical depth in the V band $\tau_V$, directly linked
with the total dust mass of the galaxy, $M_d$, through the relation

\begin{equation}
M_d= \frac{4 \pi \tau_V h_d^2}{\sigma_V},
\label{eq:tau}
\end{equation}

where $\sigma_V$ is the extinction cross section per unit mass 
for the dust in the V band.

\item
The recent star formation rate SFR, which determines the intrinsic
luminosity of the model galaxy in the UV (i.e. in the range from 912 to
3650\AA).

\item
The fraction $F$ of the locally absorbed UV, i.e. the fraction of the UV
luminosity which is produced in the dense star-formation regions and is
reprocessed locally, not being able to escape in the diffuse interstellar
medium.

\end{enumerate}

Having constructed a theoretical SED for a given galaxy, we can now 
compare it with its observed FIR spectrum, and thus determine
values for $SFR$, dust mass and $F$.

\section{The sample of galaxies}
\label{sec:sample}

\begin{figure}
\resizebox{\hsize}{!}{\includegraphics{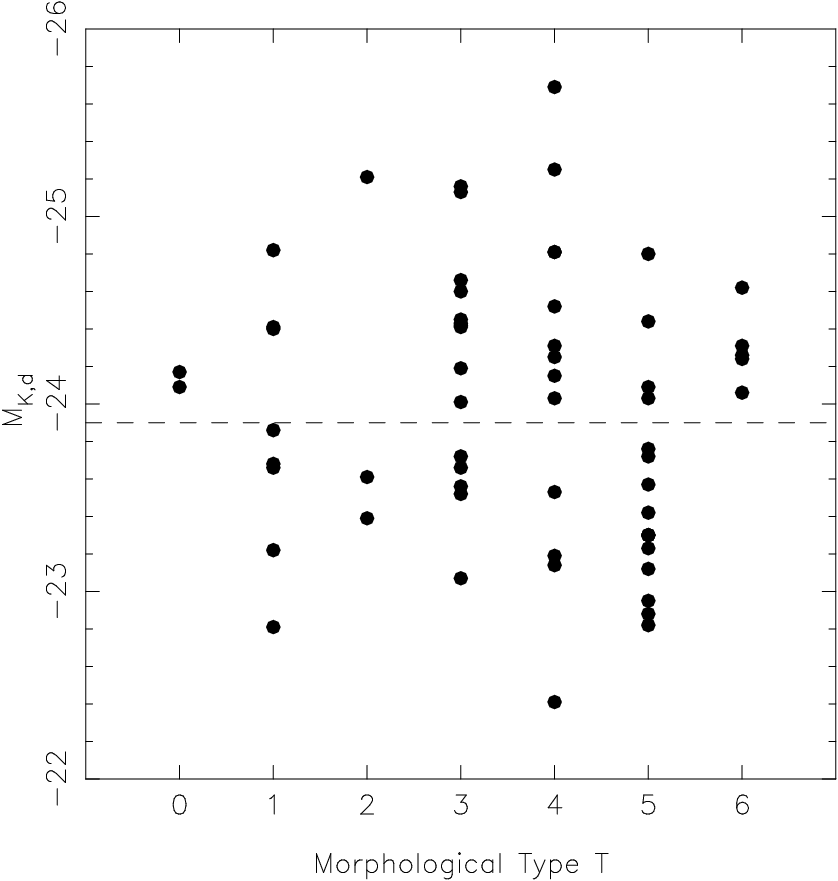}}
\caption{The disk K-band magnitudes of the galaxies in our sample plotted
as a function of galaxy morphological type, {\it T}. The dashed line
indicates the average $M_{K,d}$ value for the whole sample.}
 \label{fig:mag}
\end{figure}

Our sample of galaxies with known FIR spectra is chosen from the sample
of galaxies of DU00.  These authors present SCUBA 850 $\mathrm{\mu m}$
observations for 104 galaxies chosen from the revised IRAS Bright Galaxy
Sample (Soifer et al. \cite{soifer}). The availability of 850 $\mathrm{\mu m}$
measurements for such a large number of galaxies is the most important
property of the DU00 sample for our purposes, as measurements at the
sub-mm waveband are crucial in order to constrain the FIR emission of a
spiral galaxy. Some of these 104 galaxies have also been observed at
450 $\mathrm{\mu m}$ (Dunne et al, \cite{dunne2}).

First, we exclude galaxy pairs that were unresolved at the IRAS bands.
Since K-band magnitudes are necessary for the determination of the
geometrical characteristics of a galaxy and its
optical and near infrared intrinsic luminosity, we used 
the NASA/IPAC Extragalactic Database (NED)
to search for K-band measurements. The most recent 2MASS All-Sky Data
Release provides total $K_{S}$-band magnitudes ($M_{Ks}$) for 60 galaxies
in the DU00 sample. Note that, since the difference between the ${K_{S}}$-band
magnitudes and the K-band magnitudes of DJ96, who based their
photometry on the standards of Elias et al. (\cite{elias}), are typically
of the order of $\sim 0.02$ mags (see e.g.
www.astro.caltech.edu/~jmc/2mass/v3/transformations/), we will refer to
the 2MASS magnitudes as simply the ``K-band" magnitudes hereafter (i.e.
we assume that $M_{K}=M_{Ks}$).  We found $M_{K}$ magnitudes for 8 more
galaxies in Spignolio et al. (\cite{spin}). Apart from the availability of
$M_{K}$, the morphological type of each galaxy is also important in
determining its geometrical characteristics, as explained in the previous section.
For this reason we used the Third Reference Catalog of Bright Galaxies 
(RC3; de Vaucouleurs et al. \cite{devaucoulers}) in order to search for the $T$
value of each galaxy in the DU00 list of objects.

Our final sample consists of those 68 galaxies in the DU00 list
\cite{dunne1} with known $M_{K}$ and morphological type $0\le T \le6$.
Table~1 lists their basic properties like distance (in Mpc; column 2) as
taken from DU00, and $M_{K,d}$ (column 3). The following 3 columns list the
60 and 100 $\mathrm{\mu m}$ IRAS measurements (Soifer et al. \cite{soifer})  and the SCUBA
850 $\mu m$ measurements of DU00. Column 7 lists the logarithm
of the total gas mass, $M_{g}$.  In the cases where both atomic and
molecular hydrogen masses are available, $M_{g}$ is their sum. In the
cases where either the atomic or the molecular is available we presumed
that the total mass is twice as much. The data for the gas masses are
taken from DU00, and are based on measurements of
Sanders \& Mirabel (\cite{sanders3}),
Sanders et al. (\cite{sanders2}),
Huchtmeier \& Richter (\cite{huchtmeier}), 
Bottinelli et al. (\cite{bottinelli}), 
Sanders et al. (\cite{sanders}), 
Young et al. (\cite{young}), 
Casoli et al. (\cite{casoli}), 
Chini et al. (\cite{chini2}), 
Maiolino et al. (\cite{maiolino}), 
Solomon et al. (\cite{solomon}), 
Lavezzi \& Dickey (\cite{lavezzi}) and 
Theureau et al. (\cite{theureau}).
Note that, although there are 68 galaxies in the sample, Table~1 has 62 entries as we
can not fit well the FIR SED of 6 galaxies (see Sect.~\ref{sec:fiting}).  
As a result, we can not calculate their $SFR$, dust mass and $F$, and for
that reason these galaxies do not appear in Table~1.

In Fig.~\ref{fig:mag} we plot $M_{K,d}$ as a function of T for the
galaxies in our sample. In most cases $-25\le M_{K,d} \le-23$, with the
average $M_{K,d}$ value being equal to $-23.9$. There is no clear trend
between $M_{K,d}$ and $T$. If we compare this Figure with Fig.~6 in DJ96, we
see that, for all $T$, the $M_{K,d}$ magnitudes of the galaxies in the
present sample are brighter than the magnitudes of the
galaxies in the DJ96 sample which included all $T$. Since DJ96 uses a statistically complete
sample of spiral galaxies, we conclude that the galaxies in our
sample are more luminous than the average, typical spirals, irrespective
of their morphological type.

\section{Fitting the model to the data}
\label{sec:fiting}

\begin{figure}
\resizebox{\hsize}{!}{\includegraphics{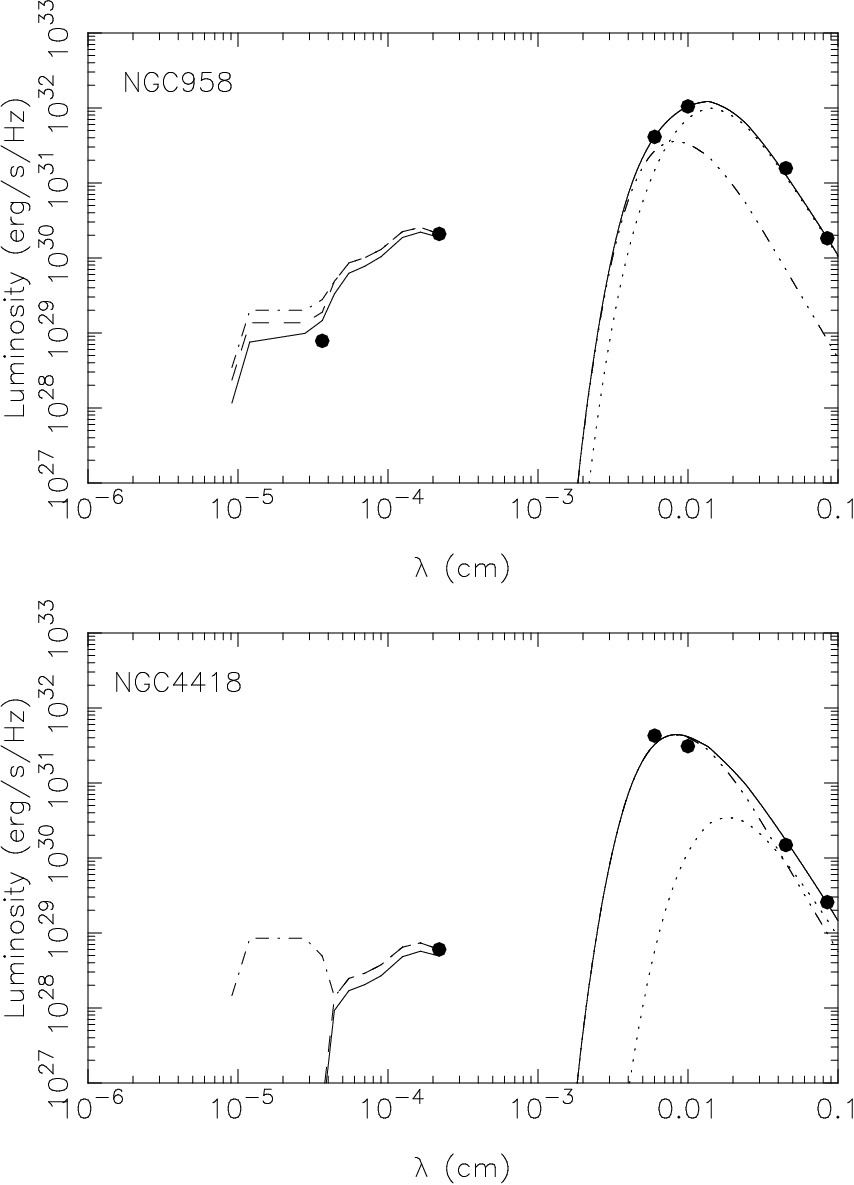}}
\caption{Examples of a good (upper plot) and a bad (bottom plot) model fit  
to the FIR spectrum of a galaxy. The three lines in the UV-optical bands
represent  the intrinsic luminosity (dot-dashed line),  the luminosity
after subtracting the locally absorbed radiation (dashed line), the luminosity
after taking into account the absorption, both locally and from the
diffuse dust (solid line). The three-dot-dashed, dotted, and solid lines in the FIR band
represent the emission from dust in the HII regions, the emission spectrum 
of the diffuse dust and the total FIR emission, respectively.}
\label{fig:goodbad}
\end{figure}

\begin{figure}
\resizebox{\hsize}{!}{\includegraphics{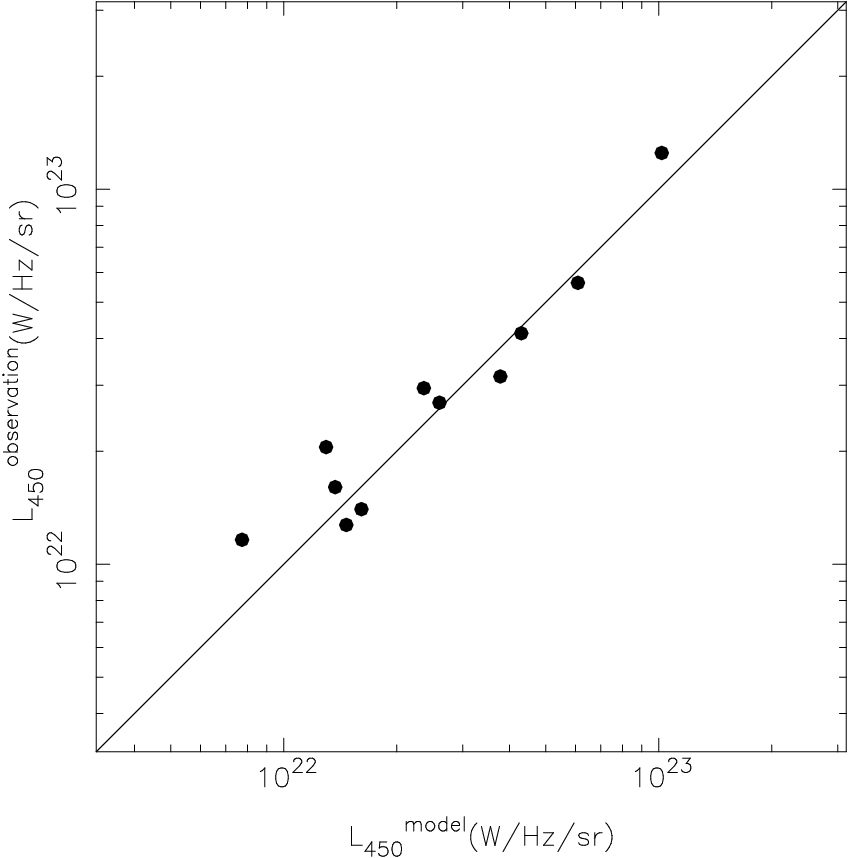}}
\caption{Comparison of the model predicted and the observed luminosity at
450 $\mu m$. The solid line shows the $L_{450}^{observed}=L_{450}^{model}$ relation, 
and {\it not} the best fitting model to the data. It is plotted in order to guide the eye, and to
demonstrate clearly the good agreement between the model predicted and
observed $L_{450}$.}
\label{fig:450test}
\end{figure}

The model SED is fitted to each set of FIR measurements by first
fixing the galaxy disk and bulge absolute magnitude in the K band to the
observed values, hence determining the geometrical parameters of the galaxy
and the luminosity of the old stellar population, as explained in
Sect.~\ref{sec:model}. Then the three free parameters of the model ($SFR$, dust mass
and $F$) are varied in order to minimize the sum of the squares of the
differences between the logarithms of the model and the observed luminosity at 60,
100 and 850 $\mathrm{\mu m}$ ($L_{60}, L_{100},$ and $L_{850}$ respectively). 
The minimization is done using the Steve Moshier C translation of the public
domain Levenberg-Marquardt solver of the Argonne National Laboratories
MINPACK mathematical library available at www.netlib.org.

Our model is able to fit very well the observed FIR spectrum of most
galaxies. As an example, we show in the upper panel of Fig.~\ref{fig:goodbad} 
the model SED (solid line) and the observed FIR spectrum of the
galaxy NGC~958. Filled circles in the right hand part of the plot show the
observed $L_{60}, L_{100}, L_{450}$ and $L_{850}$ measurements. The other
two points in  the UV and near-IR part of the spectrum show the observed
$U$- and $K$-band luminosities. The dot-dashed line in the left hand part
of the plot represents the intrinsic UV and optical SED. The dashed line
represents the UV SED after the UV absorption in the HII
regions has been subtracted.  The solid line represents the UV and optical SED
after absorption in the diffuse dust of the galaxy has also been taken
into account. In the FIR part of the spectrum, the three-dot-dashed line shows
the emitted spectrum of the HII regions, and the dotted
line shows the emitted spectrum of the diffuse dust. Finally, the
solid line represents the sum of these two components (i.e. the total
FIR emission). This figure shows clearly that the model FIR SED fits very
well the observed FIR spectrum. The predicted UV flux agrees well with the
observed U flux as well, as their ratio is smaller than two.

However, we can not fit well the observed FIR spectrum of six galaxies in our sample,  
namely, MCG+02-04-025, NGC~1614, NGC~4418, IC~860, NGC~7714 and MRK~331. 

The ratio $L_{60}/L_{100}$ in these galaxies is higher than 0.8.  
This is the maximum flux ratio value that can be explained by the 
fixed template we use for the dust emission from HII regions.  
This is the main reason for the failure of the model to provide
a good fit to the FIR spectra of these galaxies.  
In the bottom panel of Fig.~\ref{fig:goodbad},
we show a plot of the observed FIR spectrum and the best fitting model
SED in the case of NGC~4418.  The lines drawn in this plot have the same
meaning as the lines shown in the upper panel of the same figure. Although
the model fits well the SCUBA measurements, it fails to fit well the IRAS
fluxes. Comparison of the observed FIR SEDs between the two galaxies
shown in Fig.~\ref{fig:goodbad} reveals clearly the difference in the IRAS
colors of NGC~958 and NGC~4418, with the second being significantly
``warmer". In all cases where we cannot fit well the FIR SED, the best
fitting $F$ values are almost equal to 1, which in effect implies that
almost all the FIR emission in these galaxies is produced in HII
regions. Although this may well be the case, the bad quality of the model
fitting to their FIR spectra prevents us from drawing conclusive results
in these cases. While we could get a better model fit to the FIR spectra
of these six galaxies using a higher temperature for the dust emission in
the HII regions, this would lead to the adoption of an extra free
parameter to characterize the emission properties of the HII regions.
Since a detailed description of the properties of HII
regions in spiral galaxies does not lie within the scope of this paper we
decided to exclude these galaxies from the discussion hereafter. As a
result, we end up with 62 galaxies for which we have managed to fit well
their FIR spectra and hence calculate the dust mass, the $SFR$ and the $F$ for them. 
The best model fitting values are listed in Table 1, columns 8, 9, and 10
respectively.

Finally, although there are SCUBA 450 $\mathrm{\mu m}$ measurements for some of the
galaxies in our sample (Dunne et al. \cite{dunne2}), we do not take them
into account during the model fitting of their FIR spectrum. Having found
though the best fitting model parameters for each galaxy, we can now
compare the model predicted and observed flux values at 450 $\mathrm{\mu m}$.
Fig.~\ref{fig:450test} shows this comparison plot. The model predicted
flux values agree very well with the observed values. This is an important
result, as this excellent agreement justifies the dust emission properties
we have used and strongly suggests that the dust emissivity index $\beta$
is very close to $\beta=2$, as argued by Weingartner \& Draine
(\cite{weingartner}).

\section{Results}
\label{sec:results}

\input{table_data.tex}

\subsection{SFR estimates}

\begin{figure}
\resizebox{\hsize}{!}{\includegraphics{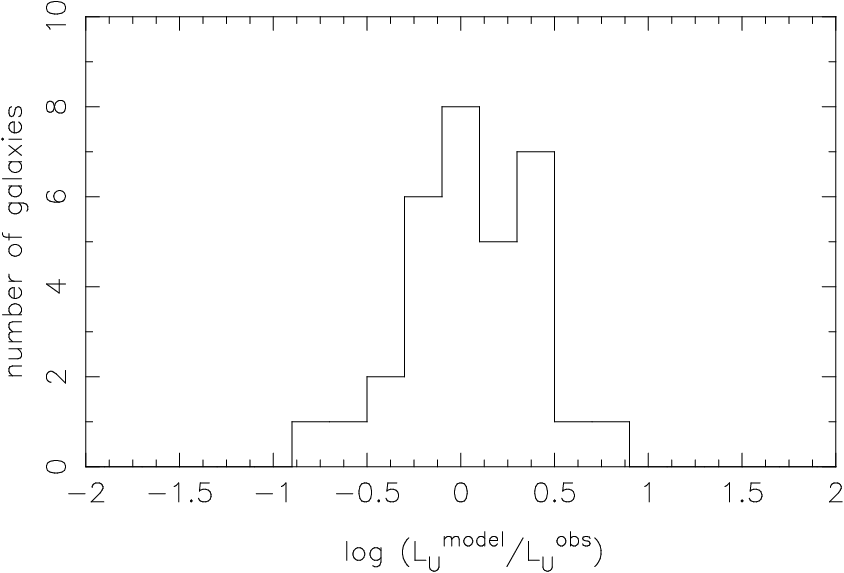}}
\caption{Comparison of the model predicted and the observed UV luminosity
for the galaxies of our sample with published photometry in the U band.}
\label{fig:uvtest}
\end{figure}

\begin{figure}
\resizebox{\hsize}{!}{\includegraphics{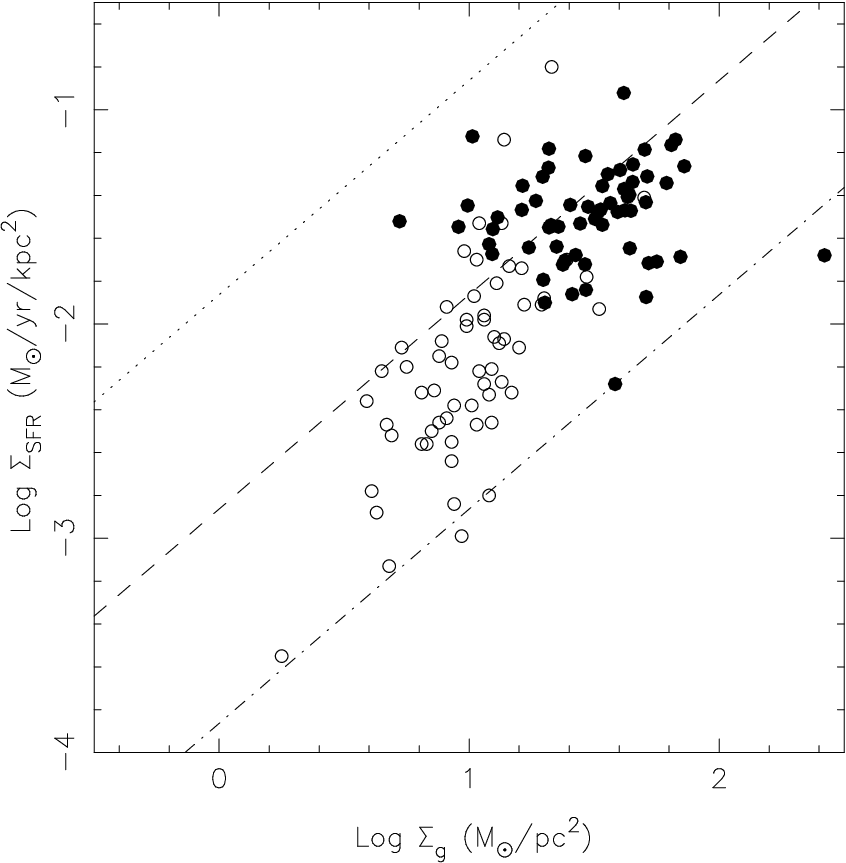}}
\caption{Star-formation-rate surface density as a function of gas surface
density for the galaxies in our sample (solid circles).  Open circles show
data from 61 normal spiral galaxies by Kennicutt (\cite{kennicutt1}). The
dotted, dashed and dot-dashed lines correspond to star formation
efficiencies of 100\%, 10\%, and 1\% in $10^8$ yr.}
\label{fig:kenni}
\end{figure}

\begin{figure}
\resizebox{\hsize}{!}{\includegraphics{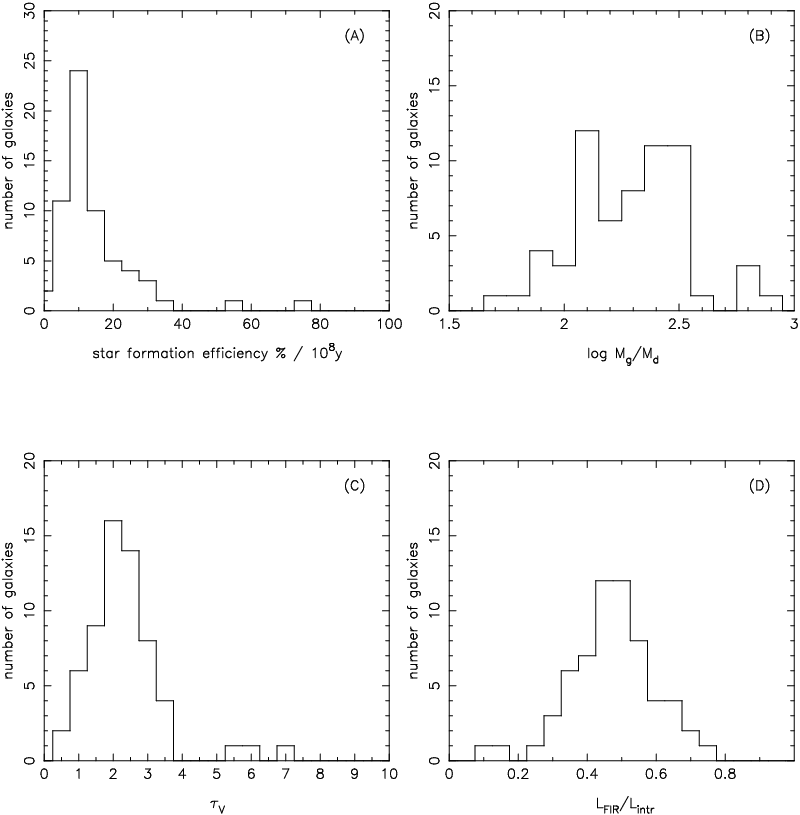}}
\caption{Sample distribution plots of the following model derived
parameters: Star formation efficiency for a time scale of $10^{8}$ years
(Panel A), gas-to-dust mass ratio (Panel B), central face-on optical depth
in the $V$ band (Panel C), FIR to total intrinsic luminosity ratio (Panel D).}
\label{fig:histo}
\end{figure}

The best fitting SFR values of the galaxies in our sample range roughly
from 2$\mathrm{M_\odot\, yr^{-1}}$ to 80$\mathrm{M_\odot\, yr^{-1}}$, 
with an average value of 19 $\mathrm{M_\odot\, yr^{-1}}$. 
These values lie between the SFR estimates for late type,
gas-rich spirals and optically-selected starburst galaxies (Kennicutt
\cite{kennicutt2}). In order to investigate whether the model computed
SFRs are realistic or not, we perform the following test. Using the RC3
catalog, we find that 32 out of the 62 galaxies in the sample have known
U-band magnitudes.  Since the model intrinsic UV emission is uniquely
determined by the adopted SFR value, the comparison between the model
predicted and the observed U-band luminosity can give us direct
information on the accuracy of the model resulting SFR.  
Fig.~\ref{fig:uvtest} shows a plot of the statistics for the ratio of the
model ($L_U^{model}$) over the observed luminosity ($L_U^{obs})$ in the
U-band. The average $L_U^{model}/L_U^{obs}$ value of 1.4 implies a good
agreement between the model and the observed luminosities.  The scatter of
the $L_U^{model}/L_U^{obs}$ values around unity is expected because a) the
true radii of the galaxies are expected to ``scatter" around the adopted
values (which were not measured directly but were based on the observed
$M_{K,d}$ values), b) the temperature of the dust in the HII regions
in the individual galaxies is naturally expected to ``scatter" around the
assumed value of $35\degr\mathrm{K}$, and c) $L_U^{model}$ is the integrated
intensity for the total $4\pi$ solid angle around the galaxy, without
taking into account the inclination of these galaxies. We conclude that
the close agreement between $L_U^{model}$ and $L_U^{obs}$ implies that the
model derived SFR values are close (within a factor of $\sim 2$) to the
intrinsic values for each galaxy in the sample. 
This fact, together with the fact that the model SEDs agree very well 
with the observed FIR spectra give us confidence that the model derived 
dust mass and $F$ values should also represent
correctly the respective intrinsic values for the galaxies under study.

The large range in SFR values that we derive reflects the corresponding
large range of absolute luminosities in the galaxies. Indeed, we find a
positive correlation between SFR and $M_{K,d}$ in the sense that more
luminous galaxies do show a larger SFR (in absolute units). However, when
we divide the SFR values by the $K$-band, disk luminosity, $L_{K,d}$, we
find that the normalized SFR values decrease with increasing luminosity,
according to the relation $SFR_{norm} \propto L_{K,d}^{-0.15}$. This
relation implies an anticorrelation between SFR per unit luminosity and
galaxy mass (as parametrized by $L_{K,d}$); more massive galaxies appear
to have a smaller SFR per unit mass. This is qualitatively similar to the
results from previous works.  For example, Gavazzi et al. (\cite{gavazzi})
also find an anticorrelation between star formation rate (normalized to
the H-band luminosity) and the mass of the galaxy (as parametrized by
the H-band luminosity).

We also investigate the relation of SFR with the gas content of the
galaxies. For this reason we divide the SFR and gas mass of
each galaxy by the surface of the disk, assuming a truncation radius of 3
disk scalelengths (Pohlen et al. \cite{pohlen}).  Fig.~\ref{fig:kenni}
shows the plot of SFR surface density as a function of gas surface density
(solid circles). In the same figure we also plot the SFR and gas surface
density measurements of 61 normal spiral galaxies from Kennicutt
(\cite{kennicutt1}). The galaxies of the present sample have larger
average gas and SFR surface densities, when compared to the respective
values of the sample of normal galaxies. This is not unexpected, as they
are IRAS bright galaxies, so one expects them to have both larger SFR
and gas densities when compared to normal galaxies. Nevertheless, the
relation between SFR surface density and gas surface density is in agreement with the same
relation as defined by the data of the normal galaxies alone.
Fig.~\ref{fig:kenni} also shows the constant, disk-averaged 1\%, 10\%, and
100\% star formation efficiency per $10^8$ years (dot-dashed, dashed, and
dotted lines respectively). The star formation efficiency is defined as
the percentage of the gas mass that is converted to stars assuming
constant SFR for a period of $10^8$ years. The distribution of the 
star-formation efficiency values is shown in panel (A) of Fig.~\ref{fig:histo}.
For almost all galaxies, the star formation efficiency lies between
5\% and 30\%.  The average value for the galaxies in our sample is 14\%,
slightly larger than the respective value of the normal spiral galaxies,
but close to the typical value for gas rich spirals.

\subsection{The dust mass estimates}

In most studies, the dust emission in spiral galaxies is modeled with one
or more gray-body components of constant temperature. Usually a ``warm''
and a ``cold'' component are assumed, their combined spectra are fitted to
the observed FIR spectra, and as a result the dust masses and the temperatures
of the components are derived (e.g. 
Alton et al. \cite{alton1}; \cite{alton2}; 
Haas et al. \cite{haas};
Davies et al. \cite{davies}; 
Lisenfeld et al. \cite{lisenfeld1}; \cite{lisenfeld2}; 
Trewhella et al. \cite{trewhella}; 
Irwin et al. \cite{irwin};
Popescu et al. \cite{popescu4};
Bendo et al. \cite{bendo}; 
B\"{o}ttner et al. \cite{bottner}; 
Dupac et al. \cite{dupac};
Hippelein et al. \cite{hippelein}).

While our model also assumes two components, they are introduced in a
natural way and they correspond to physically distinct dust components
within the galaxy. The temperature of the ``warm"  component is kept fixed
at $35\degr\mathrm{K}$, according to the results from FIR observations of dense,
HII regions, but the temperature of the ``cold" component is not
fixed. In fact, this component does not even have the same temperature
throughout its volume. Instead, its temperature varies with the distance
from the galaxy's center, and is computed through the solution of the
radiative transfer equation. As a result, our dust mass calculations are
less affected by assumptions about the dust temperature of the warm
component and the wavelength dependence of the dust emissivity.

The dust masses that we derive range from $M_d\sim 10^7\mathrm{M_\odot}$ to
$M_d\sim 3\times10^8\mathrm{M_\odot}$ and lie within the accepted range for spiral
galaxies. 
The large range of dust-mass values reflect again the large
range of absolute magnitudes.  Knowing the gas mass as well, we can now
compute the gas-to-dust mass ratio ($M_{g}/M_{d}$) for each galaxy in the
sample. The distribution of the $M_{g}/M_{d}$ values is shown in panel (B)
of Fig.~\ref{fig:histo}. The value of this ratio lies within 100 and 300
for most of the galaxies, while its mean value is 250, in excellent
agreement with the generally accepted value for our galaxy. In the past,
gas-to-dust ratios of the order of $\sim 5-10$ times higher than the
Galactic value were found for bright IRAS galaxies (Devereux \& Young
\cite{devereux}; Sanders et al. \cite{sanders}). Our results demonstrate
that when the full band FIR spectrum is taken into account and modeled
with a realistic physical model, then the gas-to-dust mass ratio for bright
IRAS galaxies is similar to the Galactic value.

Apart from the dust mass, we can also calculate the central face-on
optical depth, using equation (1).  Panel (C) in Fig.~\ref{fig:histo}
shows the distribution of the central face-on optical depth at the V
band, $\tau_V$. We find no correlation between $\tau_V$ and size of
the galaxy (as parametrized with the use of $M_{K,d}$ values). Most of
the galaxies exhibit a central face-on optical depth between 1 and 4, while
the mean value is 2.3. In other words, we find that these bright IRAS
galaxies tend to have an average central face-on optical depth somewhat
larger than unity.

\subsection{The F estimates}

The value of the fraction $F$ of the locally absorbed UV, covers the full
range from almost zero to almost one for the galaxies in our sample. In
most cases though this value lies between 0.3 and 0.6. We find no
correlation between $F$ and either SFR or $M_{K,d}$. However, we do find a
strong correlation between $F$ and the ratio of the luminosity 
at 60$\mathrm{\mu m}$ over the luminosity at 100 $\mathrm{\mu m}$: $F\propto (L_{60}/L_{100})^{2.3}$.
 In other words, we find that warmer FIR spectra 
(in terms of the 60 over 100 $\mathrm{\mu m}$ flux ratios) 
exhibit higher percentage of local absorption of UV in star forming regions.

Finally, we also compute the ratio of the FIR luminosity over the total
intrinsic UV/optical/NIR luminosity, $L_{FIR}/L_{intr}$. The distribution
of the ratio values is shown in panel (D) of Fig.~\ref{fig:histo}. The
values cover the range from 0.4 to 0.6, with the mean value being 0.54.
Popescu \& Tuffs (\cite{popescu2}) studied 28 Virgo cluster galaxies
ranging from S0 to Sm and found $L_{FIR}/L_{intr}$ values between 0.15 and
0.5.  Our sample seems to be complementary to theirs in the sense that our
sample consists of large, gas rich spirals that are strong IRAS sources,
while their sample consists of normal galaxies. Yet we find no
correlation of $L_{FIR}/L_{intr}$ with morphological type as they do.

We also find no correlation between $L_{FIR}/L_{intr}$ and either SFR or
$M_{K,d}$.  However, we do find a significant correlation between
$L_{FIR}/L_{intr}$ and $F$, in the sense that as $F$ increases, then
$L_{FIR}/L_{intr}$ increases as well, following the relation
$L_{FIR}/L_{intr}=0.25+0.5 F$. Perhaps then, the difference in the
average $L_{FIR}/L_{intr}$ value between normal spirals and the bright
IRAS galaxies of the present sample may simply reflect the difference in
their average $F$ values.

The fraction of the locally absorbed UV is directly linked with
the overall attenuation in the UV band (due to both locall absorption and
absorption from the diffuse dust). To enable the comparison with
other works we have also calculated the extinction of each galaxy
at 1800\AA. The calculated values are given in the last locumn of Table~1.
The extinction at 1800\AA ranges from 0.2 to 3.1 magnitudes

\section{FIR diagnostic relations}
\label{sec:diagnostics}

Since our model provides a good fit to the observed FIR spectra of the
galaxies in the sample, we now proceed to investigate if we can find
significant correlations between their observed FIR properties and the SFR
and dust-mass values.  To this aim, we use the results listed in Table
1 and correlate the model derived parameter values with various observed
FIR quantities. The set of ``observed quantities" vs ``model parameters",  
which provide the most significant correlations (or alternatively relations
with the least possible scatter in then) can then be used as a basis to
derive diagnostic relations between SFR, dust mass and FIR observations.

\subsection{SFR diagnostics}

\begin{figure}
\resizebox{\hsize}{!}{\includegraphics{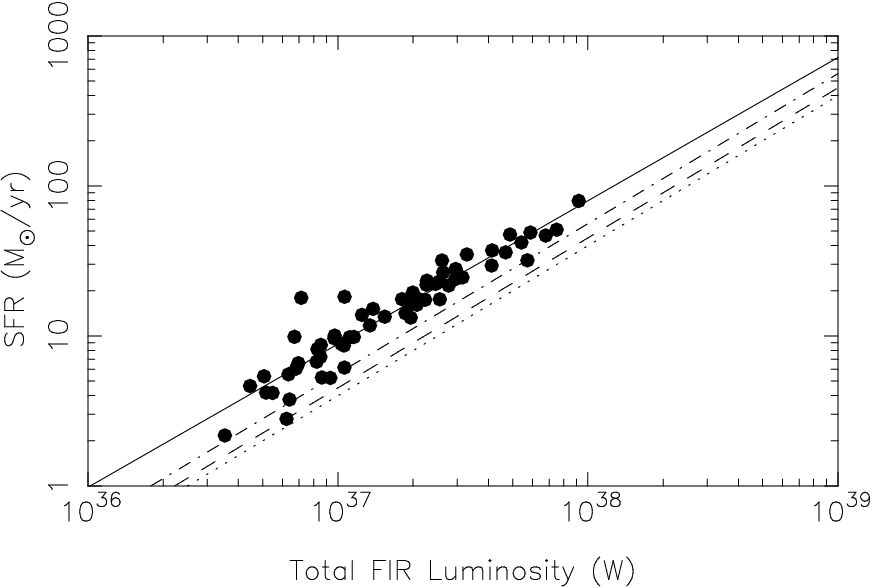}}
\caption{Star formation rate as a function of Total FIR luminosity.
The solid line is a best linear fit (in log-log space).
The dashed line is drawn according to the $L_{FIR}-$SFR relation of
Kennicutt (\cite{kennicutt2}) for starburst galaxies. The dot-dashed
line represents the same relation as calibrated by Buat \& Xu (\cite{buat}).
Finally the  dotted line represents the calibration of Panuzzo et al. (\cite{panuzzo}).}

\label{fig:sfr1}
\end{figure}

\begin{figure}
\resizebox{\hsize}{!}{\includegraphics{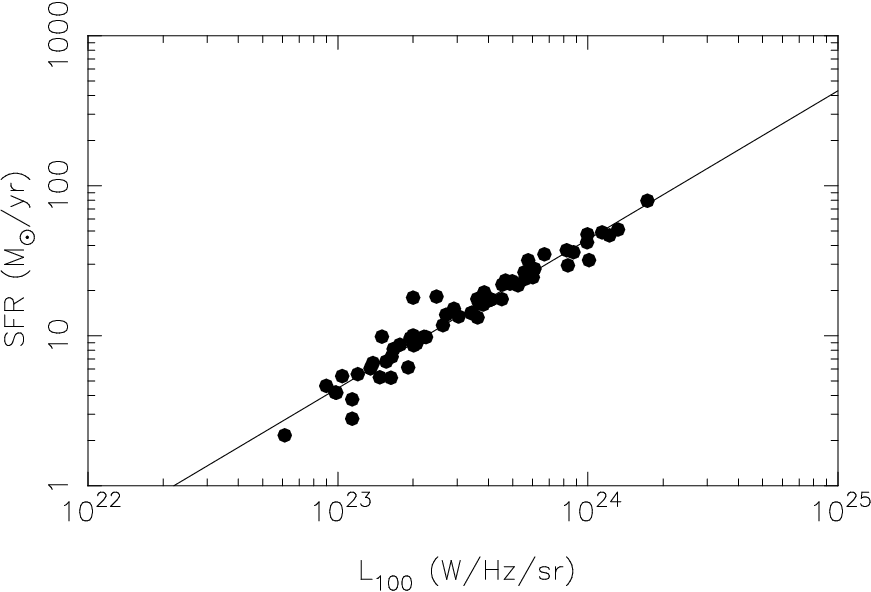}}
\caption{Star formation rate ploted as a function of $L_{100}$.
The solid line is a best linear fit (in log-log space).}
\label{fig:sfr2}
\end{figure}

We find that the SFR is well correlated with the total FIR luminosity,
$L_{FIR}$ (computed using the model FIR spectrum over the wavelength range
10 - 2000 $\mathrm{\mu m}$), and $L_{60},L_{100}, L_{850}$. The best correlations,
i.e. the ones with the least scatter, are found between SFR and $L_{FIR}$ and
between SFR and $L_{100}$ 
(shown as solid lines in Fig.~\ref{fig:sfr1} and Fig.~\ref{fig:sfr2}, respectively). 
The best linear-model fit to the SFR vs
$L_{FIR}$ plot (in log-log space) yields the following relation:

\begin{equation}
\log\left(\frac{SFR}{\mathrm{M_\odot\,yr^{-1}}}\right)=-34.4 + 0.95 \log\left(\frac{L_{FIR}}{\mathrm{W}}\right).
\label{eq:sfr1}
\end{equation}

The relation given by Buat \& Xu (\cite{buat})
is drawn also in Fig.~\ref{fig:sfr1} as a dot-dashed line
\footnote{In their paper,Buat \& Xu (\cite{buat}) used the FIR from 40 to 120 $\mathrm{\mu m}$.
In our spectra roughly 70\% of the total FIR emission lies in this range. To compare
our result with theirs we have corrected their relation accordingly}.
These authors quote a margin of error of the order of 2. On average, the
scatter in our case is smaller, as can be seen from
Fig.~\ref{fig:sfr1}. However, as Fig.~\ref{fig:uvtest} shows, we do expect
our SFRs to be uncertain by a factor of 2, so the
uncertainty of the SFR of equation (2) should be of that order.
The dotted line represents  the relation between FIR and SFR from
Panuzzo et al. (\cite{panuzzo}). The large disagreement can be attributed to
the higher attenuation of their model which causes the absorption of more stellar
radiation.
Similar SFR$-L_{FIR}$ relations have been introduced in the past for
starburst galaxies. The relation with the calibration of Kennicutt
(\cite{kennicutt2}) is drawn in Fig.~\ref{fig:sfr1} as a
dashed line. Bell (\cite{bell2}) presents a similar result. It is clear
that, in the case of our sample, a given FIR luminosity yields
roughly twice as much SFR as starburst galaxies due to the reduced
percentage of starlight which is absorbed and re-emitted in the FIR in the
latest.

The very good correlation between SFR and $L_{100}$ shown in
Fig.~\ref{fig:sfr2} is not surprising. The maximum of the
FIR emission lies between 100 and 200 $\mathrm{\mu m}$ and therefor $L_{100}$
traces well the total FIR emission, which in turn depends on SFR. The best
linear model fit to the SFR vs $L_{100}$ plot (in log-log space) yields
the following relation:

\begin{equation}
\log \left( \frac{SFR}{\mathrm{M_\odot\,yr^{-1}}}\right)=-22.1 + 0.99 \log\left(\frac{L_{100}}{\mathrm{W\,sr^{-1}\,Hz^{-1}}}\right).
\label{eq:sfr2}
\end{equation}

Note that this relation may be more useful in practice than the relation
defined by equation~(2), as $L_{100}$ is a directly measured quantity,
while $L_{FIR}$ in equation~(\ref{eq:sfr1}) is derived from the model FIR
SED. In practice, empirical methods will introduce an extra component of
uncertainty in the estimation of $L_{FIR}$ and hence SFR.

It is not clear whether equations~(\ref{eq:sfr1}) and (\ref{eq:sfr2}) can
be used to estimate the SFR for {\it all} spiral galaxies with
morphological type index $0\le T\le 6$, similar to the galaxies studied in
this work. This depends on whether the contribution of the old stellar
population to dust heating changes with galaxy luminosity or not. 
The average value of the ratio of dust heating caused by the old and the young
stellar population is 0.2 for the galaxies in our sample, indicating, that
80\% of the dust heating originates in the young stellar population.
We find no correlation between this ratio and $M_{K}$, i.e. the galaxy's
luminosity. 
However, the galaxies studied in this work have $M_{K}$
magnitudes which are brighter than these of the typical spiral galaxies. 
We caution then that equations~(\ref{eq:sfr1}) and (\ref{eq:sfr2}) may not be
applicable to less luminous galaxies, i.e. galaxies with $M_K>-23$.  This
issue will be resolved in the future, when SCUBA observations of a large
number of spiral galaxies with magnitudes fainter than, say, $M_K=-23$ will
be available. In this case, we will be able to study their FIR spectrum in
a similar fashion, and investigate whether equations~(\ref{eq:sfr1}) and
(\ref{eq:sfr2}) can be applied to them as well.

\subsection{Dust mass diagnostics}

\begin{figure}
\resizebox{\hsize}{!}{\includegraphics{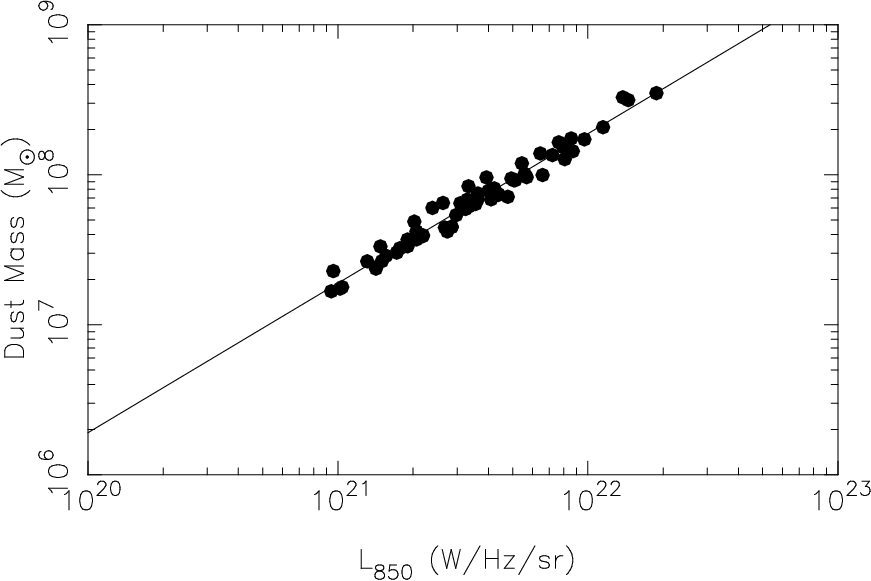}}
\caption{Dust mass as a function of Luminosity at 850 $\mathrm{\mu m}$.
The solid line shows the best linear fit (in log-log space). }
\label{fig:dustmass}
\end{figure}

We find that $M_{d}$ is well correlated with $L_{850}$.
Fig.~\ref{fig:dustmass} shows the
diffuse dust mass plotted as a function of  $L_{850}$. 
This figure shows clearly the excellent correlation
between $M_{d}$ and $L_{850}$.  In fact, the relation between these
two quantities is linear although we do not
consider a single value for the dust temperature of the ``cold" component
(in which case the relationship between $M_{d}$ and
$L_{850}$ is linear by identity). The best-fitting linear model (in
log-log space) yields the relation,

\begin{equation}
log\left(\frac{M_d}{\mathrm{M_\odot}}\right)= -13.6 + 0.997 \times \left(\frac{L_{850}} {\mathrm{W\,sr^{-1}\,Hz^{-1}}}\right).
\end{equation}

While this relation can be used to derive the dust mass from the $850\mu
m$ flux one has to keep in mind that the calibration depends on the
adopted cross section per unit mass $\sigma_{850}$ of the dust at the 850
$\mu m$. The use of a larger value for $\sigma_{850}$ will result in a
reduced $M_d$ value and vice versa.  In this study we have implicity used the value
of 0.505 $\mathrm{cm^2\,gr^{-1}}$, as suggested by the model of Weingartner \& Draine
(\cite{weingartner}). However, other theoretical models suggest values that
go up to 3 $\mathrm{cm^2\,gr^{-1}}$ (e.g. Mathis \& Whiffen \cite{mathis}; Ossenkopf \&
Henning \cite{ossenkopf}).  Furthermore, there are observational attempts
to measure $\sigma_{850}$ mainly by comparing the dust attenuation in the
optical band with the emission at 850 $\mathrm{\mu m}$. Alton et al.  
(\cite{alton3}) cite a value of 1.29 $\mathrm{cm^2\,gr^{-1}}$ (for NGC~891), 
Bianchi et al. (\cite{bianchi2}) based on data from the Galaxy argue for 0.37
$\mathrm{cm^2\,gr^{-1}}$, while in a subsequent work based on Barnard 68, 
Bianchi et al. (\cite{bianchi3}) find a higher value of 1.5 $\mathrm{cm^2\,gr^{-1}}$. 
Finally, James et al. (\cite{james}) use a different method based on
metalicity and find $\sigma_{850}$ equal to 0.7 $\mathrm{cm^2\,gr^{-1}}$.

\section{Summary}
\label{sec:summary}

We use the realistic physical model of P00 in order to study the FIR spectrum
of 62 bright, IRAS galaxies. The model takes into account the 3D spatial
distributions of both the old and young stellar populations, and the
geometry of the dust. The intrinsic optical and NIR SEDs of each galaxy are
determined by the observed K-band magnitude, while the UV SED is
proportional to the star-formation rate through the use of the population
synthesis model of Fioc \& Rocca-Volmerange (\cite{fioc}). By solving the
radiative transfer equation, we derive the FIR SED which we then fit to
the observed FIR spectra of each galaxy in order to derive values for
the SFR, the dust mass and the fraction, F, of the UV radiation which is 
locally absorbed in the HII regions.

The sample of galaxies studied in this work consists of late-type spiral
galaxies ($0\le T\le 6$) whose $M_{K}$ magnitudes are brighter than the
average K-band magnitudes of normal spiral galaxies with the same
morphology. Apart from their brighter K-band magnitudes, we also find that
their average SFR rate, star-formation efficiency, SFR and gas surface
densities are larger than the respective values for normal spiral galaxies.
At the same time though, the densities of the SFR and the gas follow the same
relation as the one defined by normal galaxies. The SFR per unit
luminosity decreases with the galaxy's luminosity, in agreement with a
similar trend seen in normal galaxies. The galaxies under study are bright
IRAS galaxies not just because they are intrinsically more luminous, but
also because, on average, they have larger $L_{FIR}/L_{intr}$ ratio values
than normal galaxies. Our results show also a correlation between
$L_{FIR}/L_{intr}$ and $F$. This result suggests that the larger than
``usual" $L_{FIR}/L_{intr}$ values are due to the fact that $F$, i.e. the
fraction of the UV radiation which is reprocessed locally in HII
regions, is larger than that in typical spirals. It is not clear though
what the reason for this is, as we find no correlation between F and SFR
(in absolute units), or $M_{K}$, or even surface densities of SFR and gas within
the galaxies in the present sample.

Our dust-mass values range from $10^{7} M_{\odot}$ to $3\times 10^8
M_{\odot}$. We find that the average gas-to-dust mass ratio of the galaxies in
our sample is almost identical to the Galactic value, and that the average
central face-on optical depth in the $V$ band is 2.3, i.e. somewhat larger than
unity. This result implies that the innermost parts of these galaxies
would not be transparent in the optical part of the spectrum if they were
seen face-on. This large optical depth though may not be
representative of the whole spiral-galaxy population. Our results suggest
that the gas surface density correlates positively with the dust surface
density. Since these galaxies have a gas surface density which, on
average, is larger than the gas surface density of normal spiral galaxies,
we expect the same trend to hold for their dust surface density, and hence
for $\tau_{V}$ as well. Furthermore, the optical depth we derived depends
on the dust mass which in turn depends on the adopted dust cross section
per unit mass, as we discussed in Section 6.2. If $\sigma_{850}$ is indeed
larger than the adopted value of 0.505 $\mathrm{cm^2\,gr^{-1}}$, then the
computed $\tau_{V}$ values should decrease proportionally.

Finally, using the model derived SFR and dust-mass estimates, we find
diagnostic relations between SFR and $L_{FIR}$ or $L_{100}$, and between
$M_{d}$ and $L_{850}$. We believe that the SFR vs $L_{FIR}$ or $L_{100}$
relations can be used to estimate (within a factor of $\sim 2$) the SFR of
a galaxy with morphological type $0\le T \le 6$ and $M_{K} \le -23$. As
for the $M_{d}$ vs $L_{850}$ relation, the uncertainty of the derived
$M_{d}$ values depends on the uncertainty associated with the cross
section per unit mass of the dust at 850 $\mathrm{\mu m}$.

\begin{acknowledgements}
We wish to thank C. Popescu and R. Tuffs for taking the time
to read this manuscript and give us their comments.

This publication makes use of data products from the Two Micron All Sky
Survey, which is a joint project of the University of Massachusetts and
the Infrared Processing and Analysis Center/California Institute of
Technology, funded by the National Aeronautics and Space Administration
and the National Science Foundation.

This research has made use of the NASA/IPAC Extragalactic Database (NED)
which is operated by the Jet Propulsion Laboratory, California Institute
of Technology, under contract with the National Aeronautics and Space
Administration.

Finally we would like to thank the referee V. Buat for her useful
comments and suggestions.

\end{acknowledgements}

\end{document}

%% file: table_data.tex
\begin{table*}
\centering
\caption[]{Data and model parameters of the galaxies in our sample. Details in Section \ref{sec:sample}.}
\begin{tabular}{lllllllllll}
\hline
NAME & distance & $M_{K,d}$ & log $L_{60} $ & log $L_{100}$ & log $L_{850}$ & log $M_g$ &  log $M_d$ & log SFR & $F$ & $A_{1800}$ \\ 
- & Mpc & - & W/Hz/sr & W/Hz/sr & W/Hz/sr & $M_{\odot}$ & $M_{\odot}$ & $M_{\odot}/yr$ & - & - \\ 
\hline
NGC 23        & 61.0  & -24.82 & 23.49 & 23.72 & 21.70 & 10.22 & 7.96 & 1.34 & 0.53 & 1.12 \\ 
UGC 556       & 62.0  & -23.52 & 23.29 & 23.56 & 21.45 & 9.93  & 7.65 & 1.25 & 0.39 & 0.90 \\ 
NGC 470       & 32.0  & -23.56 & 22.83 & 23.06 & 21.31 & 9.78  & 7.69 & 0.58 & 0.69 & 1.62 \\ 
UGC 903       & 34.0  & -23.14 & 22.93 & 23.19 & 21.28 & 9.82  & 7.57 & 0.83 & 0.49 & 1.12 \\ 
NGC 520       & 30.0  & -23.86 & 23.44 & 23.61 & 21.45 & 10.13 & 7.65 & 1.24 & 0.65 & 1.45 \\ 
NGC 697       & 42.0  & -24.44 & 22.96 & 23.43 & 21.56 & 10.67 & 7.84 & 1.14 & 0.16 & 0.50 \\ 
NGC 772       & 33.0  & -25.13 & 22.84 & 23.39 & 21.47 & 10.57 & 7.73 & 1.26 & 0.03 & 0.21 \\ 
NGC 877       & 52.0  & -24.52 & 23.48 & 23.78 & 21.93 & 10.48 & 8.24 & 1.39 & 0.47 & 1.11 \\ 
NGC 958       & 77.0  & -25.69 & 23.51 & 23.92 & 22.16 & 10.74 & 8.50 & 1.47 & 0.29 & 0.72 \\ 
NGC 992       & 55.0  & -23.76 & 23.50 & 23.66 & 21.63 & 10.27 & 7.91 & 1.24 & 0.76 & 2.00 \\ 
NGC 1134      & 49.0  & -24.45 & 23.31 & 23.56 & 21.74 & 10.58 & 8.08 & 1.12 & 0.58 & 1.32 \\ 
UGC 2403      & 55.0  & -23.68 & 23.34 & 23.54 & 21.51 & 10.14 & 7.77 & 1.15 & 0.63 & 1.46 \\ 
UGC 2982      & 71.0  & -24.26 & 23.62 & 23.92 & 21.92 & 10.46 & 8.17 & 1.57 & 0.41 & 1.00 \\ 
NGC 1667      & 61.0  & -24.80 & 23.34 & 23.76 & 21.76 & 10.18 & 7.98 & 1.50 & 0.16 & 0.49 \\ 
NGC 2782      & 34.0  & -23.66 & 23.03 & 23.21 & 21.42 & 9.81  & 7.81 & 0.72 & 0.81 & 2.23 \\ 
NGC 2856      & 35.0  & -23.30 & 22.86 & 23.08 & 21.02 & 9.45  & 7.25 & 0.74 & 0.50 & 1.01 \\ 
NGC 2966      & 27.0  & -22.41 & 22.61 & 22.79 & 20.98 & 9.44  & 7.36 & 0.34 & 0.77 & 2.04 \\ 
NGC 2990      & 41.0  & -22.88 & 22.95 & 23.21 & 21.25 & 9.99  & 7.51 & 0.86 & 0.47 & 1.11 \\ 
IC 563        & 81.0  & -23.61 & 23.26 & 23.56 & 21.81 & 10.10 & 8.14 & 1.22 & 0.44 & 1.10 \\ 
IC 564        & 80.0  & -24.31 & 23.18 & 23.60 & 21.88 & 10.10 & 8.22 & 1.24 & 0.29 & 0.82 \\ 
NGC 3110      & 67.0  & -24.41 & 23.70 & 24.00 & 21.91 & 10.64 & 8.10 & 1.68 & 0.36 & 0.89 \\ 
NGC 3221      & 55.0  & -24.62 & 23.33 & 23.75 & 21.86 & 10.49 & 8.13 & 1.42 & 0.23 & 0.66 \\ 
NGC 3367      & 40.0  & -24.09 & 22.98 & 23.29 & 21.31 & 10.07 & 7.57 & 0.98 & 0.33 & 0.69 \\ 
NGC 3583      & 28.0  & -23.72 & 22.74 & 23.18 & 21.15 & 9.90  & 7.37 & 0.99 & 0.12 & 0.38 \\ 
MCG+00-29-023 & 102.0 & -24.42 & 23.73 & 23.94 & 21.92 & 10.68 & 8.17 & 1.56 & 0.60 & 1.41 \\ 
UGC 6436      & 137.0 & -24.43 & 24.00 & 24.24 & 22.27 & 10.44 & 8.54 & 1.90 & 0.50 & 1.22 \\ 
NGC 3994      & 41.0  & -23.23 & 22.91 & 23.22 & 21.24 & 10.05 & 7.48 & 0.91 & 0.35 & 0.80 \\ 
NGC 4045      & 26.0  & -23.22 & 22.63 & 22.95 & 20.97 & 9.43  & 7.22 & 0.67 & 0.32 & 0.67 \\ 
NGC 4273      & 32.0  & -23.30 & 23.00 & 23.30 & 21.49 & 9.94  & 7.81 & 0.94 & 0.45 & 1.09 \\ 
NGC 4433      & 40.0  & -23.39 & 23.33 & 23.53 & 21.52 & 10.06 & 7.79 & 1.15 & 0.63 & 1.50 \\ 
NGC 4793      & 33.0  & -23.57 & 23.11 & 23.46 & 21.43 & 10.04 & 7.65 & 1.18 & 0.26 & 0.68 \\ 
NGC 5020      & 45.0  & -24.03 & 23.01 & 23.28 & 21.59 & 10.37 & 7.98 & 0.79 & 0.63 & 1.49 \\ 
NGC 5104      & 74.0  & -24.41 & 23.55 & 23.83 & 21.68 & 10.13 & 7.85 & 1.54 & 0.33 & 0.75 \\ 
NGC 5256      & 111.0 & -25.21 & 23.93 & 24.09 & 21.99 & 10.66 & 8.24 & 1.67 & 0.74 & 1.77 \\ 
UGC 8739      & 67.0  & -24.81 & 23.40 & 23.79 & 21.90 & 10.42 & 8.18 & 1.45 & 0.28 & 0.72 \\ 
NGC 5371      & 34.0  & -24.81 & 22.77 & 23.30 & 21.18 & 10.35 & 7.39 & 1.36 & 0.01 & 0.16 \\ 
NGC 5426      & 35.0  & -23.12 & 22.55 & 22.99 & 21.52 & 10.72 & 7.92 & 0.62 & 0.32 & 0.88 \\ 
NGC 5427      & 36.0  & -24.03 & 22.94 & 23.31 & 21.52 & 10.38 & 7.83 & 0.95 & 0.33 & 0.78 \\ 
NGC 5600      & 31.0  & -22.95 & 22.69 & 23.02 & 21.01 & 9.35  & 7.24 & 0.73 & 0.30 & 0.69 \\ 
NGC 5653      & 47.0  & -24.66 & 23.34 & 23.67 & 21.64 & 9.87  & 7.87 & 1.37 & 0.30 & 0.68 \\ 
NGC 5665      & 30.0  & -22.82 & 22.71 & 22.99 & 21.12 & 9.27  & 7.42 & 0.62 & 0.48 & 1.08 \\ 
NGC 5676      & 28.0  & -24.15 & 22.96 & 23.35 & 21.56 & 10.18 & 7.88 & 0.99 & 0.30 & 0.73 \\ 
NGC 5713      & 25.0  & -23.53 & 23.10 & 23.34 & 21.34 & 10.11 & 7.60 & 0.99 & 0.49 & 1.08 \\ 
NGC 5792      & 26.0  & -24.19 & 22.77 & 23.06 & 21.38 & 10.31 & 7.78 & 0.45 & 0.67 & 1.53 \\ 
NGC 5900      & 33.0  & -23.07 & 22.89 & 23.25 & 21.28 & 9.93  & 7.52 & 0.94 & 0.30 & 0.76 \\ 
NGC 5936      & 53.0  & -24.01 & 23.36 & 23.66 & 21.61 & 9.95  & 7.84 & 1.34 & 0.37 & 0.85 \\ 
NGC 5937      & 37.0  & -23.66 & 23.12 & 23.42 & 21.50 & 9.51  & 7.78 & 1.07 & 0.41 & 0.96 \\ 
NGC 5962      & 26.0  & -23.42 & 22.77 & 23.13 & 21.31 & 9.77  & 7.62 & 0.78 & 0.34 & 0.80 \\ 
NGC 5990      & 51.0  & -24.40 & 23.36 & 23.59 & 21.44 & 9.53  & 7.62 & 1.29 & 0.43 & 0.85 \\ 
NGC 6052      & 63.0  & -23.30 & 23.38 & 23.58 & 21.55 & 10.18 & 7.81 & 1.21 & 0.62 & 1.49 \\ 
NGC 6181      & 32.0  & -23.72 & 22.95 & 23.30 & 21.34 & 10.07 & 7.59 & 1.00 & 0.29 & 0.68 \\ 
NGC 7448      & 29.0  & -23.19 & 22.83 & 23.14 & 21.19 & 9.83  & 7.46 & 0.82 & 0.37 & 0.83 \\ 
ZW 453.062    & 100.0 & -24.24 & 23.83 & 24.00 & 21.82 & 9.76  & 8.00 & 1.62 & 0.66 & 1.56 \\ 
NGC 7541      & 36.0  & -24.25 & 23.39 & 23.69 & 21.71 & 10.39 & 7.96 & 1.35 & 0.40 & 0.92 \\ 
ZW 475.056    & 109.0 & -24.60 & 24.00 & 24.12 & 21.94 & 10.21 & 8.16 & 1.71 & 0.83 & 2.29 \\ 
NGC 7591      & 66.0  & -24.31 & 23.51 & 23.75 & 21.75 & 10.42 & 8.01 & 1.38 & 0.53 & 1.18 \\ 
NGC 7674      & 116.0 & -25.25 & 23.83 & 24.00 & 22.14 & 10.79 & 8.52 & 1.50 & 0.83 & 2.39 \\ 
NGC 7678      & 47.0  & -24.06 & 23.16 & 23.48 & 21.60 & 10.00 & 7.89 & 1.13 & 0.38 & 0.89 \\ 
NGC 7771      & 57.0  & -25.16 & 23.80 & 24.06 & 22.06 & 10.38 & 8.32 & 1.69 & 0.47 & 1.05 \\ 
UGC 1351      & 61.0  & -24.09 & 23.33 & 23.61 & 21.69 & 10.09 & 7.98 & 1.24 & 0.46 & 1.08 \\
UGC 1451      & 66.0  & -24.17 & 23.44 & 23.70 & 21.64 & 9.93  & 7.87 & 1.36 & 0.44 & 0.98 \\
NGC 3094      & 32.0  & -22.81 & 23.05 & 23.17 & 21.17 & 9.65  & 7.52 & 0.72 & 0.91 & 3.07 \\
\end{tabular}
\end{table*}